\documentclass{birkjour} 

\usepackage[T1]{fontenc}    
\usepackage{hyperref}       
\usepackage{url}            
\usepackage{booktabs}       
\usepackage{amsfonts}       
\usepackage{nicefrac}       
\usepackage{microtype}      
\usepackage{multirow}
\usepackage{amsmath}
\usepackage[caption=false]{subfig}

\usepackage{graphicx}

\usepackage[noadjust]{cite}

\usepackage[dvipsnames]{xcolor}
\newcommand{\RRR}{\mathbb{R}^{3}}
\newcommand{\CGA}{\mathbb{R}_{4,1}}

\usepackage{algorithm,algcompatible,amsmath}
\algnewcommand\INPUT{\item[\textbf{Input:}]}%
\algnewcommand\OUTPUT{\item[\textbf{Output:}]}%

\newcommand{\ed}{\end{document}}
\newcommand{\ms}[1]{#1 \, \mathrm{ms}}
\newcommand{\rad}[1]{#1 \, \mathrm{rad}}
\renewcommand{\sec}[1]{#1 \, \mathrm{sec}}

\begin{document}

%
%
%
%
%
%
%
%
%
\title[A Geometric Algorithm for Cutting, Tearing, and Drilling Models]{An All-In-One Geometric Algorithm for Cutting, Tearing, and Drilling\\ Deformable Models} 
\author[M. Kamarianakis]{Manos Kamarianakis}

\address{%
Department of Mathematics \& Applied Mathematics, \\
University of Crete, \\
Voutes Campus, 70013 Heraklion, Greece\\
Orchid ID: 0000-0001-6577-0354}

\email{m.kamarianakis@gmail.com}

\thanks{The authors are affiliated with the University of Crete, Greece and  the ORamaVR company (\url{http://www.oramavr.com}). \\
This is an extended version of work originally presented in the CGI 
2020 conference, on the ENGAGE workshop. \cite{Kamarianakis:2020is}}

\author[G. Papagiannakis]{George Papagiannakis}
\address{Department of Computer Science, \\
University of Crete, \\
Voutes Campus, 70013 Heraklion, Greece\\
Orchid ID: 0000-0002-2977-9850}
\email{george.papagiannakis@gmail.com}

\subjclass{Primary 68U05}

\keywords{Conformal Geometric Algebra (CGA), Skinning, Interpolation, Cutting Algorithm, Tearing Algorithm, Drilling Algorithm, Keyframe Generation}

\date{\today}


\begin{abstract}
Conformal Geometric Algebra (CGA) is a framework that allows the representation of objects, such as points, planes and spheres, and deformations, such as translations, rotations and dilations as uniform vectors, called \emph{multivectors}. In this work, we demonstrate the merits of multivector usage with a novel, integrated rigged character simulation framework based on CGA. In such a framework, and for the first time, one may perform real-time cuts and tears as well as drill holes on a rigged 3D model. These operations can be performed before and/or after model animation, while maintaining deformation topology. Moreover, our framework permits generation of intermediate keyframes on-the-fly based on user input, apart from the frames provided in the model data.  We are motivated to use CGA as it is the lowest-dimension extension of dual-quaternion algebra  that amends the shortcomings of the majority of existing animation \& deformation techniques. Specifically, we no longer need to maintain objects of multiple algebras and constantly transmute between them, such as matrices, quaternions and dual-quaternions, and we can effortlessly apply dilations. Using such an all-in-one geometric framework allows for better maintenance and optimization and enables easier interpolation and application of all native deformations. Furthermore, we present these three novel algorithms in a single CGA representation which enables cutting, tearing and drilling of the input rigged model, where the output model can be further re-deformed in interactive frame rates. These close to real-time cut,tear and drill algorithms can enable a new suite of applications, especially under the scope of a medical VR simulation.
\end{abstract}

\maketitle

\section{Introduction} 
\label{sec:introduction}

In this work, we introduce a novel algorithm to perform drill in the rigged model and provide further background knowledge for representing and applying translations, rotations and dilations (uniform scalings) in multivector form. Furthermore, we give better insight regarding multivector interpolation and provide the updated performance results of our optimized cutting algorithm.

Rigged models and their animation and deformation techniques have been among the most studied topics in computer graphics since their inception, and especially in the past few years due to the rapid growth of the industry of Virtual/Augmented Reality and computer games. 

Although the linear-blend skinning algorithm for rigged models \cite{magnenat1988joint} has not radically changed over the years, the demand for more robust and efficient real-time implementations of the animation, led researchers into developing more complex mathematical frameworks to enhance the overall performance and decrease running times. Originally \cite{Alexa:2002ij}, the animation techniques were based on matrix representation of the three basic 
deformations: translation, rotation and dilation. The core idea was to be able to apply these deformations to 3D point by simply multiplying the respective matrices, in the desired order, with the homogeneous coordinates of the point. Since matrix multiplications are extremely fast to perform due to GPUs' ability of parallel processing, matrices became and still remain the favorite representation class of deformations for the majority of current state-of-the-art skeletal animation frameworks.

The major drawback of using matrices was discovered when the need of creating interpolated keyframes highlighted the fact that the interpolation result of two rotation matrices does not correspond to a rotation matrix. The idea of using the original Euler angles instead of the derived rotation matrix did not solve the problem as it yielded an even greater one; the famous \emph{gimbal lock}. Modern implementations tackle the issue, using \emph{quaternions}; an algebra of~4 dimensions, originally introduced by Hamilton in 1843. Quaternions, often denoted by $\mathbb{H}$, are an extension of the complex numbers, using two more negative dimensions, i.e., they include, besides $i$, two more distinct imaginary basis elements $j,k$ such that $j^2=k^2=-1$. It was proved that a certain subset of quaternions, called \emph{unit quaternions}, could encapsulate the essence of a rotation and also support interpolation. 

The idea of using unit quaternions to store rotations provided a solution to the matrix interpolation problem and remains until today the world standard in computer graphic's modern engines. However, it also introduced the need to constantly transmute rotations from quaternion to matrix form and vice versa in every intermediate step, adding an extra performance burden. Matrices are still needed in such implementations to store translation and dilation data, while vertices are kept in homogeneous coordinates. 

As in improvement to this situation, an algebraic extension of quaternions called 
\emph{dual quaternions} was used \cite{Kenwright:2012tl}. A specific subset of these 8-dimensional objects, called the \emph{unit dual quaternions}, was proved to be able to uphold both rotation and translation data and still allow for effortless and inexpensive linear blending. Nevertheless, this advance did not solve the uniformity problem, however it reduced artifacts appearing during animation \cite{Kavan2008}, while further post-processing can be used to further minimize them \cite{Kim:2014gb}. 

Our approach utilizes the CGA framework to perform both model animation and more complex techniques such as cutting, tearing and drilling. CGA is an algebra containing of dual-quaternions, where all entities such as vertices, spheres, planes as well as rotations, translations and dilation are uniformly expressed as \emph{multivectors} \cite{DietmarFoundations,DorstBook,Wareham:2004js}. The usage of multivectors allows model animation without the need to constantly transmute between matrices and (dual) quaternions, enabling dilation to be properly applied along with translation \cite{Papagiannakis:2013va,Papaefthymiou:2016dx}. Furthermore, the interpolation of two multivectors of the same type correctly produces the expected intermediate result \cite{Hadfield:2019cx}, which makes creation of keyframes trivial to implement. Finally, usage of the proposed framework demands a single representation type for all data and results, which is the current trend in computer graphics \cite{Muller:2016kt}.

The use of Conformal Geometric Algebra and multivector representation allows the creation of simpler algorithms to perform complicated tasks, as fundamental geometric predicates are baked in the framework. For example, the intersection of two planes can be determined by simply evaluating they geometric product. 

Therefore, complex operations such as cutting, tearing and drilling a model are now easier to be accomplished, with near real-time results. Such operations have become a major research topic as they appear in increasing frequency in real-time simulation applications, for both academic as well as industrial purposes. Current algorithms \cite{Bruyns:2002jc,CutSurvey} handle such deformations using tetrahedral mesh representations of the model, which demands a heavy 
pre-processing to be performed. Since originally introduced, cutting and tearing methods have been upgraded and extended to allow almost real-time results, using mostly finite element methods, intuitive optimization and heavy pre-processing \cite{Bielser:gh,Mor:2000jj,Bruyns:2001ki}. To make the final results even more realistic, physics engines utilizing position-based dynamics are used to simulate soft-tissue cuts at the expense of performance \cite{Bielser:1999ez,Bender2014,Berndt2017}.

\textbf{Our contribution:} The novelty of our work involves the complete implementation of rigged model animation in terms of CGA, extending the work of Papaefthymiou et al. \cite{Papaefthymiou:2016dx} in a  python-based implementation that enables keyframe generation on-the-fly. The original animation equation involving matrices is translated to its equivalent multivector form (see Section~\ref{sub:multivector_form_of_the_animation_equation}) and all information required to apply the linear blend skinning algorithm (vertices, animation data) is obtained from the model and translated as multivector. This enables us to have future animation models in CGA representation only, which, in combination with an optimized GPU multivector implementation, produces faster results under a single framework. Another major novelty of our work is the cutting, tearing and drilling algorithms that are being applied on top of the previous framework; given the input animated model, we perform cuts, tears and drills on the model surface with the ability to further re-deform the newly processed model. The subpredicates used in these algorithms utilize the multivector form of their input, so they can be implemented in a CGA-only framework. Their design was made in such a way that little to no pre-processing of the input model is required while allowing a future integration with a physics engine. Furthermore, using our method, we can  generate our own keyframes instead of just interpolating between pre-defined ones. Our all-in-one CPU python implementation is able to process an existing animation model (provided in .dae or .fbx format) and translate the existing animation in the desired CGA form while further tweaks or linear-blend deformations are available in a simple way to perform. Such an implementation is  optimal as far as rapid prototyping, teaching and future connection to deep learning is concerned. It also constitutes the base for interactive cutting, tearing and drilling presented in Section~\ref{sub:cutting_tearing_drilling}. The simplicity and robustness of 
our algorithms design promise real-time results if run in a compiled programming language such as C++ or C\#. 



\section{Introduction to Conformal Geometric Algebra} 
\label{sec:introduction_to_conformal_geometric_algebra}

The Conformal Geometric Algebra (CGA) used in this paper can be seen as another algebra containing dual-quaternions which allows round elements such as spheres to be represented as objects of this algebra, i.e., as \emph{multivectors}. To be more precise, CGA is the lowest possible extension where this is possible. Being able to represent round elements in conjunction with the ability to reflect on objects using the so-called \emph{sandwich operation} presented in the following sections, CGA is also able to represent dilators (uniform scaling) as multivectors. 
Therefore, CGA is a geometric algebra where round elements (points, spheres, circles), flat elements (lines, planes, point pairs) and all basic deformations (translations, rotations, dilations) can be expressed explicitly in multivector form. 

In order to create the model of 3D CGA, we extend the basis $\{e_{1}, e_{2},e_3\}$ of the original Euclidean space $\mathbb{R}^3$ by two elements $e_+$ and $e_-$. These elements have positive and negative signature respectively, i.e., it holds that $e_+^2=-e_-^2=1$. The resulting non-Euclidean space is usually denoted as $\mathbb{R}^{4,1}$ while the Clifford (geometric) algebra of $\mathbb{R}^{4,1}$ is denoted as $\CGA$ or $\mathcal{G}(4,1)$.

It is convenient to define a \emph{null} basis given by the original basis vectors $e_1,e_2,e_3$ of $\RRR$ and 
\begin{equation}
  e_{o} = \frac{1}{2}(e_{-} -e_{+}), \ \ \ \   e_{\infty} = e_{-}+e_{+}.
\end{equation}
\noindent
The elements $e_o$ and $e_\infty$ are called \emph{null} vectors because $e_o^2=e_\infty^2=0$, where the operation implied is the geometric product described in the following sections. 


\subsection{Vector Objects of $\CGA$}
\label{sub:vector_objects_of_CGA}

A generic vector $Y$ of $\CGA$ is a linear combination of the basis elements $ \{e_1,e_2,e_3,e_\infty,e_o\}$, i.e.,
\begin{equation}
Y = y_1e_1+y_2e_2+y_3e_3+y_{\infty}e_{\infty}+y_oe_o, \ \ \ y_i\in\mathbb{R}.  
\end{equation}

Note that CGA is a projection space where the elements $Y$ and $Z$ are equivalent if and only if there is a $\lambda\in\mathbb{R}$ such that $Y=\lambda Z$. Due to this equivalence, we usually assume, without loss of generality, that the coordinate of $e_o$ is either~$0$ or~$1$. In this algebra, points, spheres and planes are easily represented as vector objects 
of the space, as described below. 

\begin{description}
  \item[Points] A point $x = (x_1,x_2,x_3) = x_1e_1+x_2e_2+x_3e_3$ of $\mathbb{R}^{3}$ is \emph{up-projected} into the conformal vector 
    \begin{align}
    X &= x + \frac{1}{2} x^2 e_{\infty} +e_o \notag\\
		  &= x_1e_1+x_2e_2+x_3e_3 + \frac{1}{2}(x_1^2+x_2^2+x_3^2)e_{\infty}+e_o.
    \end{align}
  \item[Spheres] A sphere $s$ of the $\RRR$, centered at  $x = (x_1,x_2,x_3)$ with radius $r$ is \emph{up-projected} into the conformal vector
    \begin{align}
    S &= X - \frac{1}{2} r^2 e_{\infty} \notag\\
      &= x_1e_1+x_2e_2+x_3e_3 + \frac{1}{2}(x_1^2+x_2^2+x_3^2-r^2)e_{\infty}+e_o,
    \end{align} \noindent where $X$ is the image of $x$ in $\CGA$. 
  \item[Planes] A plane $\pi$ of the original space, with Euclidean distance $d$ from the origin,
    perpendicular to the \emph{normal} vector  $\vec{n} = (n_1,n_2,n_3)$ is \emph{up-projected} into the conformal vector 
    \begin{align}
    \Pi = \vec{n} + de_{\infty} = n_1e_1 + n_2e_2 + n_3e_3 + de_{\infty}.
    \end{align}
\end{description}



\subsection{Products in $\CGA$}
\label{sub:products_in_}

There are three major products in $\CGA$: the inner, the outer and the geometric. Each of these products is initially defined among the vectors $e_1$, $e_2$, $e_3$, $e_-$, $e_+$, $e_o$, $e_{\infty}$. The respective definition is then extended to any element (a \emph{multivector}) of the space. Below we present some of the basic properties of these products; further information 
can be found in \cite{DorstBook,DietmarFoundations}.

\begin{description}
  \item[Inner] The inner product (denoted by $\cdot$) of the basis elements is defined as follows:
    \begin{itemize}
    \item $e_i\cdot e_j:=\delta_{ij}$ for $i,j\in\{1,2,3,+\}$,
    \item $e_- \cdot e_-:=-1$,
    \item $e_- \cdot e_j:=0$ for $j\in\{1,2,3,+\}$,
    \item $e_o \cdot e_o:=e_{\infty} \cdot e_{\infty}=0$,
    \item $e_o \cdot e_{\infty}:=-1$,
    \item $e_i \cdot e_j:=0$ for $i\in\{1,2,3,+\}$ and $j\in\{o,\infty\}$.
    \end{itemize}
  \item[Outer]
The outer product of the basis elements $e_i$ and $e_j$ is denoted as $e_i\wedge e_j$. The outer product is an associative operation that can be applied to more than two elements, e.g., 
$e_i\wedge e_j\wedge e_k$ and $e_i\wedge e_j\wedge e_k \wedge e_\infty$ are properly defined. The outer product of $k$ basis vectors is called a $k$-\emph{blade} and $k$ is usually referred to as the \emph{grade} of this blade. A sum of $k$-blades is called a $k$-vector and the addition of $k$-vectors of different grades is a \emph{multivector}. 

The importance of the outer product derives from the fact that it allows us, in certain cases, to obtain the intersection of two objects by simply evaluating their outer product. Specifically, a circle (resp. line) can be seen as the intersection - outer product of two spheres (resp. planes). The outer product of a circle with an intersecting sphere or equivalently, the outer product of three intersecting spheres represent a set of two points, usually referred to as \emph{a point pair}. 

  \item[Geometric] The most important product in $\CGA$ is the so-called \emph{geometric} product. For the basis vectors $e_i$ and $e_j$, their geometric product $e_ie_j$ is defined 
  as the addition of the outer and inner product of the elements, i.e., 
	\begin{equation}
   e_ie_j := e_i\wedge e_j + e_i\cdot e_j. \nonumber
  \end{equation}   
  Note that, by the definition, $e_ie_j = e_i\wedge e_j$ for every $i,j\in\{1,2,3,\infty,o \}$ such that $i\neq j$ and $\{i,j\}\neq\{\infty,o\}$. 
\end{description}


\subsection{Dual Objects} 
\label{sub:dual_objects}
 
First, let us denote the \emph{pseudoscalar} $I$ of $\CGA$,
\begin{equation}
I:=e_1\wedge e_2\wedge e_3\wedge e_+\wedge e_- = e_1\wedge e_2\wedge e_3\wedge e_{\infty}\wedge e_o.
\end{equation}

 \noindent
 Using $I$, we may define the dual object $m^\star$ of a multivector $m$ is  to be
  \begin{equation}
  m^\star:=-mI,
  \end{equation}
 where the operation between $m$ and $I$ is the geometric product. Notice that it holds that $(m^\star)^\star = - m$ and therefore we can easily obtain the normal form~$m$ of an object from it's 
 dual form~$m^\star$ and vice versa.

 The dual form of certain objects holds strong geometric meaning, as described below. 
 \begin{itemize}
   \item The outer product of 4 non-coplanar points yields the dual 
    form of the sphere defined by these points. 
   \item The outer product of 3 non-collinear points and $e_{\infty}$ 
   yields the dual form of the plane defined by these points.
   \item The outer product of 3 non-coplanar points yields the dual 
    form of the circle defined by these points. 
   \item The outer product of 2 points and $e_{\infty}$ yields the dual 
    form of the line defined by these points. 
 \end{itemize}


\subsection{Rotations, Translations and Dilations} 
\label{sub:rotations_translations_and_dilations}

So far we have shown that objects (or their duals) such as points, planes, circles, spheres, lines and point pairs are represented as multivectors. However, the beauty and versatility of this algebra comes from its ability to also represent rotations, translations and dilations as 
multivectors as described below. 

\par{\textbf{Rotation.}} A rotation in CGA is encapsulated in a multivector
\begin{equation}
R := \exp\left(- b\frac{\phi}{2}\right) = \exp\left(- I_3u\frac{\phi}{2}\right) 
   = \cos\left(\frac{\phi}{2}\right) - uI_3\sin\left(\frac{\phi}{2}\right),
\end{equation}
where $\phi$ is the angle of the rotation, $b$ is the normalized plane of the rotation,~$u$ is the normalized axis of the rotation and  $I_3:=e_1e_2e_3$. All products are geometric products and $\exp(\cdot)$ denotes the exponential function. The inverse multivector of $R$ is 
\begin{equation}
R^{-1} := \exp\left(b\frac{\phi}{2}\right) = \exp\left(I_3u\frac{\phi}{2}\right) 
        = \cos\left(\frac{\phi}{2}\right) + uI_3\sin\left(\frac{\phi}{2}\right).
\end{equation}
\par{ \textbf{Translation.}} The multivector
\begin{equation}
T := \exp\left(- \frac{1}{2}te_{\infty}\right) = 1 - \frac{1}{2}te_{\infty},
\end{equation}
where $t = t_1e_1+t_2e_2+t_3e_3$ is a euclidean vector, represents a translation by~$t$ in CGA. The inverse multivector 
of $T$ is 
\begin{equation}
T^{-1} := \exp\left(\frac{1}{2}te_{\infty}\right) = 1 + \frac{1}{2}te_{\infty}.
\end{equation}
\par{ \textbf{Dilation.}} The multivector 
\begin{equation}
D = 1 + \frac{1-d}{1+d}e_\infty\wedge e_o
\end{equation}
corresponds to a dilation of scale factor $d>0$ with respect to the origin. The inverse of $D$ is given by the expression
\begin{equation}
D^{-1} = \frac{(1+d)^2}{4d} + \frac{d^2-1}{4d}e_\infty\wedge e_o.
\end{equation}
An interesting remark is that, for $d=0$, it holds that 
$D=1+e_\infty\wedge e_o = 1+e_+e_-$, which is clearly not 
invertible in $\CGA$ as $(e_+e_-)^2=1$.


The conformal space model allows us to apply any or multiple of 
the operations above not only to a point but also to any object $O$
that was previously defined. Let $M_i$, for $i=1,\ldots,n$, be 
either a rotation, a translation or a dilation as defined above. 
To apply the transformations 
$M_1,M_2,\ldots M_n $ (in this order), to an object $O$, we first 
define the multivector $M:= M_n M_{n-1}\cdots M_1 $, where 
all in-between products are geometric. The object 
\begin{equation}
  O':= MOM^{-1}
\end{equation}
represents the final form of $O$ after all  
transformations are applied. 

\subsection{Interpolation of Multivectors} 
\label{sub:interpolation_of_multivectors}

Interpolation of data is an essential part for Computer Graphics as it is needed in every animation procedure of a rigged model. The poses of the model with respect to time are not stored in a continuous manner but rather at discrete time-steps. If additional intermediate frames are demanded, we have to interpolate the animation data between two provided keyframes. 

As in the case of matrix (\cite{Alexa:2002ij}) or (dual) quaternion quaternion interpolation 
\cite{Kavan2008}, a blending of two multivectors can be accomplished in various ways, yielding different results \cite{Hadfield:2019cx,wareham2004applications}. Choosing a proper interpolation 
technique is not a simple task as it may depend on the model or other factors. However, two methods remain dominant in analogue with the quaternion case: the linear and the logarithmic blending.

\begin{figure}[bt!]
\centering
\subfloat[]{\includegraphics[width=0.31\textwidth]{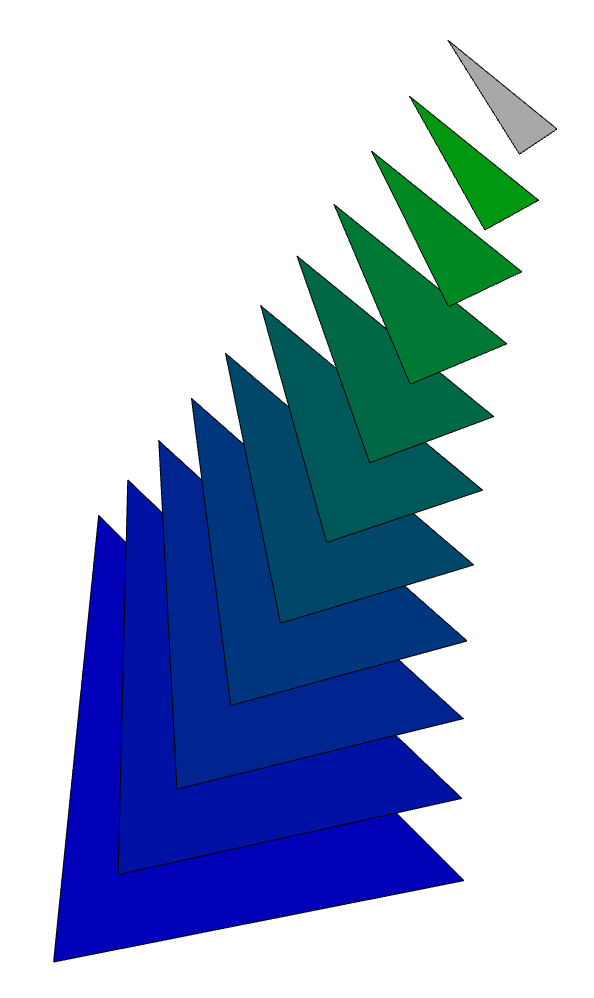}}\hspace*{0cm}
\subfloat[]{\includegraphics[width=0.31\textwidth]{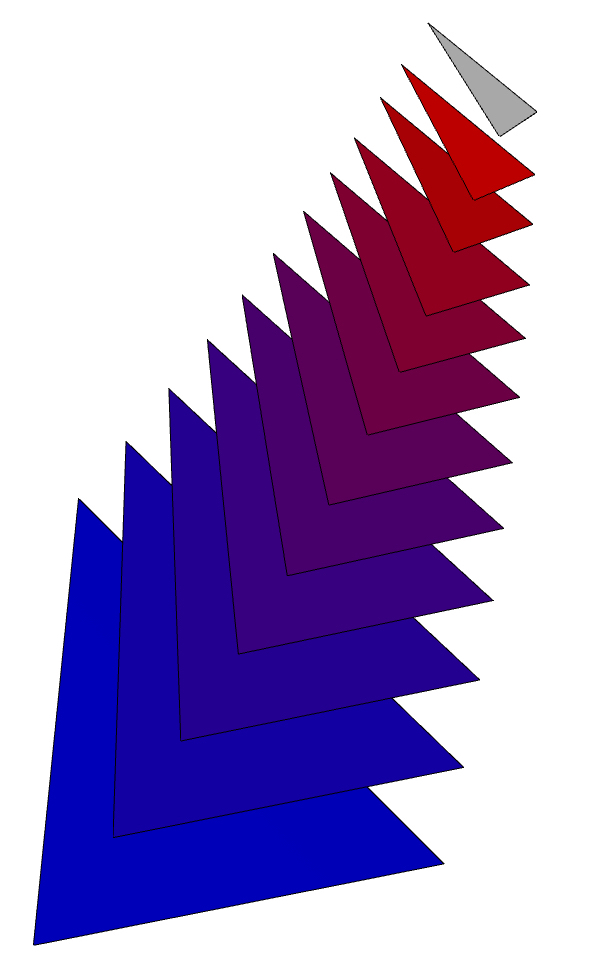}}\hspace*{0cm}
\subfloat[]{\includegraphics[width=0.31\textwidth]{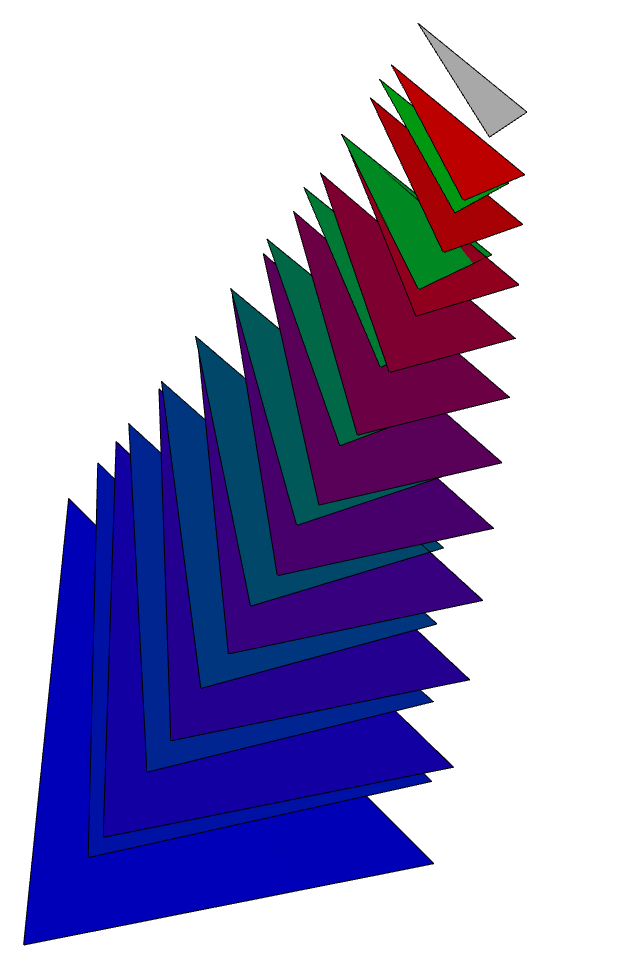}}
\caption{Linear Versus Logarithmic Interpolation. Three vertex points defining a triangular face are interpolated. This face is interpolated between the bottom blue and top grey keyframe.  Both the vertices as well as translation ($t$), rotation ($r$) and scaling($s$) data that are interpolated in their multivector form. In (A), we have used a classic linear interpolation, whereas in (B) we have used a logarithmic blending. In (C), the outcomes of these two methods are superimposed.}
\label{fig:interpolation_comparison}
\end{figure}

Linear blending of the multivectors $m_1$ and $m_2$, which, in a model animation context, may represent translations, rotations or dilations, is done by evaluating $(1-\alpha)m_1+\alpha m_2$, 
for $\alpha \in [0,1]$. Another blending method is the so-called logarithmic interpolation where we evaluate $m_1\exp(\alpha\log(m_1^{-1}m_2))$, for $\alpha \in [0,1]$, where the exponential and logarithmic function of a multivector~$m$ are approximated in our case using the respective Taylor series expansion. Notice that $m_1$ is either a rotator, a translator, a dilator with 
$d>0$ or a geometric product of such multivectors and therefore is invertible. Although not evident, one can prove that that the logarithmic interpolation method is symmetric if we interchange $m_1$ and $m_2$ as well as $\alpha$ and $1-\alpha$, by using basic exponential and logarithmic properties. Using different blendings, we obtain different results, as shown in 
Figure~\ref{fig:interpolation_comparison}. More information regarding the evaluation and properties of multivector logarithms/exponentials can be found in 
\cite{Dorst:2011,Papagiannakis:2013va,colapinto2015articulating,Wareham2007}. 

In our framework, linear blending is preferred when generating frames on-the-fly by the user, whereas logarithmic blending is used when reading the models existing animation data.

\section{State of the Art} 
\label{sec:state_of_the_art}

The current state of the art regarding skeletal model animation and deformation is based on the linear-blend skinning algorithm \cite{magnenat1988joint} and the representation of bones animation via transformation matrices and quaternions or dual-quaternions. Such an implementation allows for efficient and robust interpolation methods between keyframes. A shortcoming of such an implementation is the inability to represent a dilation as a quaternion or dual-quaternion , 
which forces the use of multiple representations and frameworks~\cite{Papaefthymiou:2016dx}. 

To be more precise regarding the mechanics of the deformation process, in the case of a simple rigged model, every bone $b_i$ amounts to an offset matrix $O_i$ and an original transformation matrix $t_i$. The skin of the model is imported as a list of vertices $v$ and a list of faces $f$. 
A bone hierarchy is also provided where $\{t_i\}$ are stored along with information regarding the animation of each joint. This information, usually referred to as \emph{TRS data}, is provided in 
the form of a quaternion, a translation vector and a scaling vector that represent respectively the rotation, displacement and scaling of the joint with respect to the parent joint for each 
keyframe (see Section~\ref{sub:state_of_the_art_representation}). 

In order to determine the position of the skin vertices at any given time~$k$ and therefore render the scene by triangulating them using the faces list, we follow the steps described below. 
Initially, a matrix $G$ is evaluated as the inverse of the transformation matrix that corresponds to the root node. Afterwards, we evaluate the \emph{global transformation matrix} for every bone $b_i$ at time $k$ and denote it as $T_{i,k}$. To evaluate all $T_{i,k}$, we recursively evaluate the matrix product $T_{j,k}t_{i,k}$ where $b_j$ is the parent bone of $b_i$, given that $T_{r,k}$ is the identity matrix (of size 4), where $b_r$ denotes the root bone. The matrix $t_{i,k}$ is a transformation matrix equal to $t_i$ if there is no animation in the model; in this case, our 
implementation allows to generate the keyframes ourselves in real-time. Otherwise, $t_{i,k}$ is evaluated as
\begin{equation}
t_{i,k} = TR_{i,k} MR_{i,k} S_{i,k}
\end{equation}
where $TR_{i,k}, MR_{i,k},S_{i,k}$ are the interpolated matrices that correspond to the translation, rotation and scaling of the bone $b_i$ at a given time $k$.

After evaluating the matrices $\{T_{i,k}\}$ for all bones $\{b_i\}$, 
we can evaluate the global position of all vertices at  
time $k$, using the \emph{rigged deformation equation}:
\begin{equation}\label{eq:animation}
V_k[m] = \displaystyle \sum_{n\in I_m} w_{m,n}G T_{n,k} O_{n}  v[m]
\end{equation}
\noindent
where
\begin{itemize}
\item $V_k[m]$ denotes the skin vertex of index $m$ (in  
homogeneous coordinates) at the animation time $k$,
\item $I_m$ contains up to four indices of bones that affect the 
vertex $v[m]$,
\item $w_{m,n}$ denotes the ``weight'', i.e., the amount of 
influence of the bone $b_n$ on the vertex $v[m]$,
\item $O_{n}$ denotes the offset matrix corresponding 
to bone $b_n$, with respect to the root bone,
\item $G$ denotes the inverse of the transformation matrix that 
corresponds to the root bone (usually equals the identity matrix) and
\item $T_{n,k}$ denotes the deformation of the bone $b_n$ at animation 
time $k$, with respect to the root bone.
\end{itemize}

\subsection{State-of-the-art Representation} 
\label{sub:state_of_the_art_representation}

The modern way to represent the TRS data of a keyframe is 
to use matrices for the translation and dilation data as well as 
quaternions for the rotation data. Let  $\{TR_i,R_i,S_i\}$, 
denote such data at keyframe $i\in\{1,2\}$, where:
\begin{itemize}
  \item $TR_i=\begin{bmatrix}
    1 & 0 & 0 & x_i\\
    0 & 1 & 0 & y_i\\
    0 & 0 & 1 & z_i\\
    0 & 0 & 0 & 1\\
  \end{bmatrix}$ and 
  $S_i=\begin{bmatrix}
    sx_i & 0 & 0 & 0\\
    0 & sy_i & 0 & 0\\
    0 & 0 & sz_i & 0\\
    0 & 0 & 0 & 1\\
  \end{bmatrix}$ represent the translation 
  by $(x_i,y_i,z_i)$ and the scale by $(sx_i,sy_i,sz_i)$ respectively, and
  \item $R_i$ is a quaternion representing the rotation.
\end{itemize} 
Note that these matrices and quaternions are extracted from 
a provided animated rigged model file (usually a *.dae 
or *.fbx file) or could be created on-the-fly by the user. 
Before quaternions, Euler angles and the derived rotation matrices 
were used to represent rotation data. However the usage of such 
matrices induced 
a great problem: a weighted average of such matrices does not correspond 
to a rotation matrix and therefore interpolating between two states would 
require interpolating the Euler angles and re-generate the corresponding 
matrix. This in turn would sometimes lead to a gimbal lock or to `candy-wrapper' artifacts such as the ones presented in \cite{Kavan2008}. 

The usage of quaternions allowed for easier interpolation techniques 
while eradicating such problems. Nevertheless, a transformation 
of the interpolated quaternion to corresponding rotation matrix 
was introduced since the GPU currently handles only matrix multiplications 
in a sufficient way for skinning reasons. Therefore, the interpolation between the two keyframes mentioned above follows the following pattern:
\begin{enumerate}
\item the matrices $TR_a = (1-a)TR_1+aTR_2$ and $S_a = (1-a)S_1+aS_2$ 
are evaluated for a given $a\in [0,1]$,
\item the quaternion $R_a = (1-a)R_1+aR_2$ is determined and finally,
\item the rotation matrix $MR_a$ that corresponds to $R_a$ is 
calculated. 
\end{enumerate}

The interpolated data $TR_a, MR_a$ and $S_a$ are then imported to the GPU
in order to determine the  intermediate frame, based on the equation\eqref{eq:animation}. The calculation of the intermediate keyframes amongst multiple ones, is either performed via explicit selection of those contained in the offline animation file or they are generated  procedurally via  interpolation blending, via the tweening method.

Using the method proposed in this paper, all data are represented in multivector form. A major implication of this change is that the interpolation between two states is done in a more clear and uniform way as presented in Section~\ref{sec:our_methodology}. This also makes the need to constantly transform a quaternion to a rotation matrix redundant, although we are now obliged to perform multivector additions and multiplications as well as down project points 
from~$\mathbb{R}^{4,1}$ to~$\mathbb{R}^{3}$ to parse them to the GPU. However, since all our data and intermediate results are in the same multivector form, we could (ideally) program the GPU
to implement such operations and therefore  greatly improve performance. 


\begin{figure}[bt!]
\centering
\subfloat[]{\includegraphics[width=0.24\textwidth]{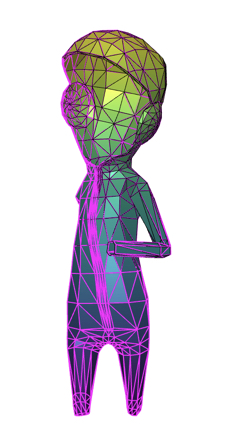}}\hspace*{0cm}
\subfloat[]{\includegraphics[width=0.31\textwidth]{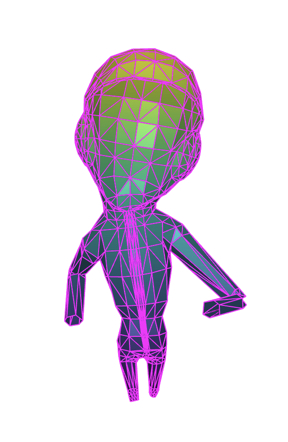}}\hspace*{0cm}
\subfloat[]{\includegraphics[width=0.24\textwidth]{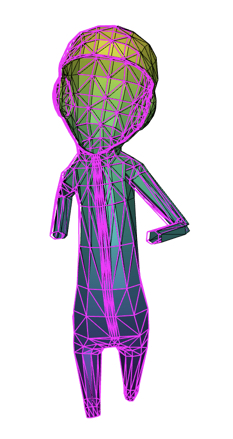}}
\caption{Skinning via the multivectors versus skinning via the dual quaternions. The original model is deformed using multivectors and depicted in magenta wireframe, superimposed with the color-graded result (based on the $z$ coordinate of each vertex) of the quaternion method for the same deformation. It is qualitatively verified that linear blending of multivectors produces similar results with the current state-of-the-art method. Evaluating the vector differences
of all vertices for the two methods, we have evaluated the approximation error assuming the quaternion method to be the correct, using the infinity ($\ell_\infty$) norm. (A) We applied a slight rotation on the neck joint, resulting in approximation error $0.3\%$. (B) We applied a slight dilation on the neck joint, approximation error is $0.00035\%$. (C) We applied a slight translation on the neck joint, approximation error $1\%$. The model used contains 1261 vertices and 1118 faces.}
\label{fig:comparison}
\end{figure}

\section{Our Algorithms and Results} 
\label{sec:our_methodology}


\subsection{Multivector Form of the Rigged Deformation Equation} 
\label{sub:multivector_form_of_the_animation_equation}

The deformation equation \eqref{eq:animation}, core of the animation algorithm, yields fast results (especially when combined with a GPU implementation) but denies us a robust way to 
dilate with respect to a bone. Our motivation is to extend and apply the animation equation for multivector input as proposed in \cite{Papaefthymiou:2016dx}. 

To be more specific regarding our method, we propose the replacement of all matrices appearing in \eqref{eq:animation} with multivectors for animation purposes. The transformation matrix of $t_i$ of each bone $b_i$ as well as all information regarding translation and rotation for each keyframe, initially extracted from the provided model file, can be easily converted to multivectors \cite{DietmarFoundations,DorstBook}. Consequently, we can evaluate the multivector $M_{i,k}$ which is equivalent to the matrix $T_{i,k}$ by following the same procedure of determining the latter (described in Section~\ref{sec:state_of_the_art}) while substituting all involved matrices with the corresponding multivectors. 

Note that various techniques can be used to interpolate between two keyframes to obtain $M_{i,k}$; 
for existing keyframes logarithmic blending is preferred \cite{Hadfield:2019cx,Kavan2008}, whereas for keyframe generation we use linear blending. In both scenarios, the intermediate results are multivectors of the correct type.

Furthermore, each offset matrix $O_n$ and each skin vertex $v[m]$ is translated to their CGA form
$B_n$ and $c[m]$ respectively. Finally, $G$ matrix is normalized to identity and is omitted in the final equation. 

Our final task is to translate in CGA terms the matrix product 
$$
T_{n,k} O_n v[m],
$$
where  apparently each multiplication sequentially applies a deformation to vertex $v[m]$. To apply the respective deformations, encapsulated by $M_{n,k}$ and $B_n$, to CGA vertex $c[m]$, we have to evaluate the \emph{sandwich geometric product} $(M_{n,k}B_{n})c[m](M_{n,k}B_{n})^\star$ where 
$V^\star$ denotes the \emph{inverse} multivector of $V$ (see \cite{Kenwright:2012tl,DietmarFoundations} for details). 

Summarizing, if the multivector form of the vertex $V_k[m]$, which corresponds to the final position of the $m$-th vertex at animation time $k$, is denoted by $C_k[m]$, then the \emph{multivector deformation equation} becomes

\begin{equation}\label{eq:cga_formula}
C_k[m] = \displaystyle 
\sum_{n\in I_m} w_{m,n}(M_{n,k}B_{n})c[m](M_{n,k}B_{n})^\star
\end{equation}
After the evaluation of $C_k[m]$ for all $m$, we can down-project 
all these conformal points to the respective euclidean ones in order to 
represent/visualize them and obtain the final result of the 
keyframe at time $k$.

The replacement of matrices with multivectors enables the introduction 
of dilations in a simple way. The multivector $M_{i,k}$ that represents
a rotation and translation with respect to the parent bone of $b_i$ 
can be replaced with $M_{i,k}D_{i,k}$ where $D_{i,k}$ is the 
corresponding dilator and the operation between them is the geometric
product. The dilator corresponds to a scale factor with respect to the 
parent bone, information that could not be easily interpreted via matrices. 
However, since the application of a motor and/or a dilator to a vertex 
is a sandwich operation, such a dilation becomes possible when 
using multivectors. 

A comparison between the results of our proposed method and the 
current state-of-the-art is shown in Figure~\ref{fig:comparison},
where we successfully apply dilation to different bones 
and obtain similar results. Rotations, dilations and translations 
are obtained in our method using multivectors only, 
under a single framework with simpler notation/implementation;
linear blending is used to interpolate between keyframes.


\subsection{Cutting, Tearing and Drilling Algorithms} 
\label{sub:cutting_tearing_drilling}

A novelty we present in this paper is the cutting, tearing and drilling
algorithms on skinned triangulated models. 
As the name suggests, the first module enables the user to 
make a planar cut of the model whereas the second is used to perform 
smaller intersections on the skin. 
The last module can be utilized to drill holes in the skinned model.
In the following sections, we provide a detailed presentation of the 
algorithms involved as well as certain implementation details.

\begin{figure}[htb]
\centering
\subfloat[]{\includegraphics[width=0.26\textwidth]{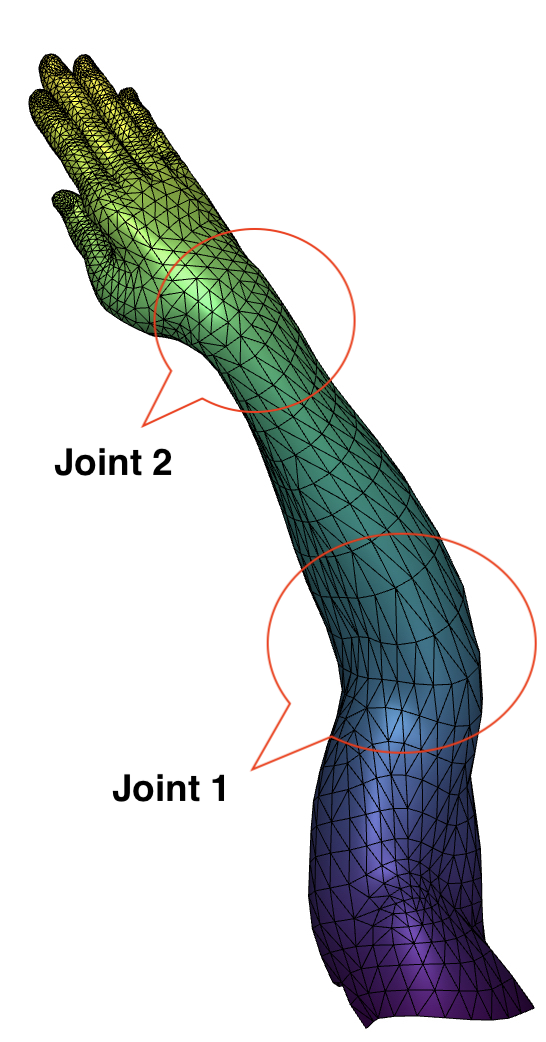}}\hspace*{5mm}
\subfloat[]{\includegraphics[width=0.25\textwidth]{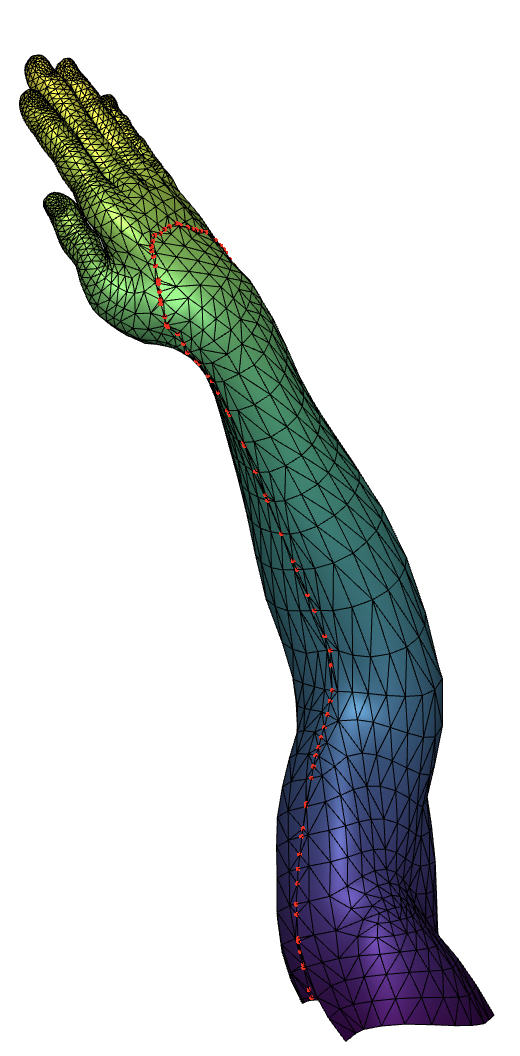}}\hspace*{0mm}\\
\subfloat[]{\includegraphics[width=0.25\textwidth]{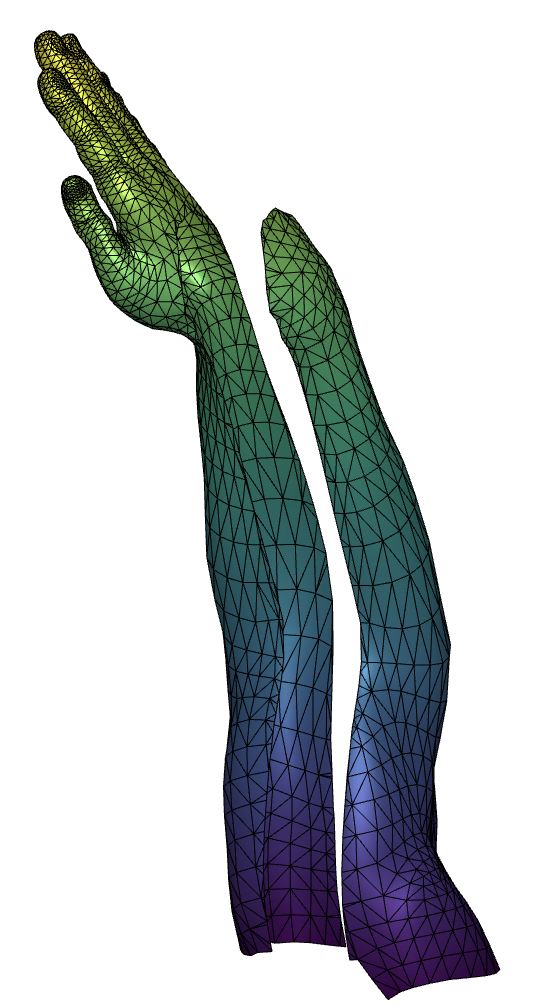}}\hspace*{5mm}
\subfloat[]{\includegraphics[width=0.33\textwidth]{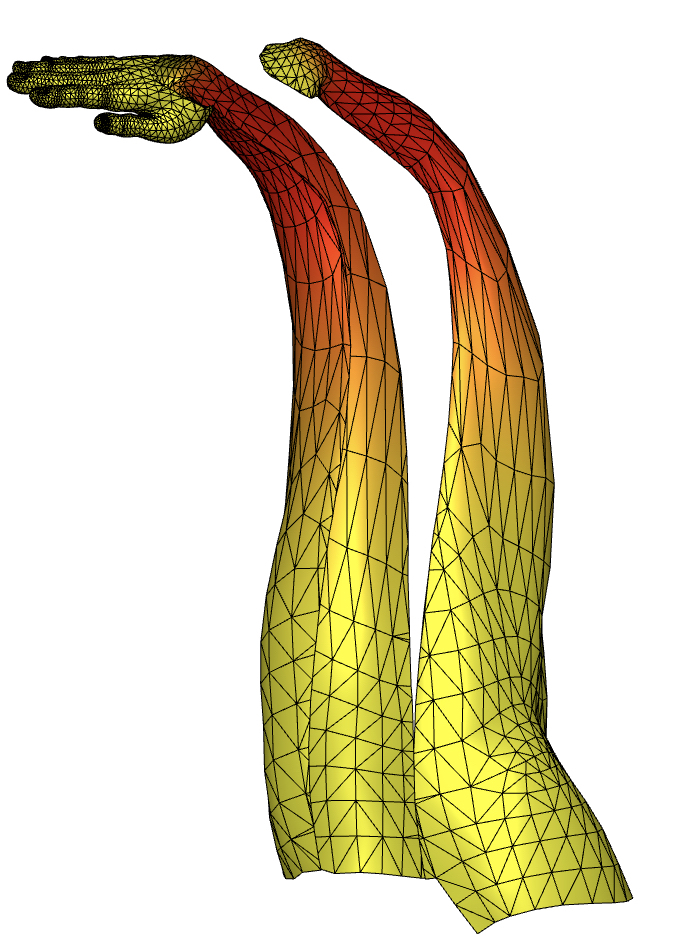}}
\caption{Cutting module intermediate steps. 
(A) The original animated model. (B) The model where the (red) intersection points of the cutting plane and the mesh are calculated and re-triangulated. (C) The model after the cut. (D) The 
model is deformed by a rotation (axis=$(0,1,1)$, $\rad{0.7}$), a translation (vector=$(13,0,0)$) and a dilation (factor = 0.5) at joint 1 (elbow), as well as another rotation (axis=$(0,1,1)$, 
$\rad{0.3}$) at joint 2 (wrist). Note that minimal artifacts occur in the final result. The vertices in (D) are colored depending on the influence of joint 1 which is mostly deformed. The vertices in (A)-(C) are colored based on their~$z$ coordinate.} 
\label{fig:cutting_module}
\end{figure}

\subsubsection{Cutting Algorithm} 
Cutting a skinned model is implemented in current bibliography in many
forms \cite{Bruyns:2002jc,ye2014improved,wang2005cutting,ji2006easy,ye2011research}. 
The most common technique is via the usage of
tetrahedral meshes \cite{Bielser:gh} which require a 
heavy pre-processing on the model and  currently do not enable 
further animation of the model or scale to VR environments. 
Our work includes an algorithm for planar model cut, where 
the final mesh is deformable, as we implemented a 
function to calculate weights for all additional vertices that 
did not originally exist (see Figure~\ref{fig:cutting_module}). 
Most of the subpredicates used in the cutting algorithm are
implemented in terms of conformal geometry and therefore can be 
used even if the model is provided in multivector form. 

Our proposed planar cut implementation is summarized as 
Algorithm~\ref{alg:cutting}. A description of how we tackle 
the weight evaluation in step~\ref{step:weights_cutting}
is found in Section~\ref{sub:implementation_details}. Our algorithm 
does not require tetrahedral meshed models and requires minimum to none 
pre-processing. It is GA-ready and the low number
of operations it demands make it suitable for VR implementations.

\begin{algorithm}[hb]
  \caption{Cutting Algorithm}
  \label{alg:cutting}
  \begin{algorithmic}[1]
    \INPUT Triangulated Mesh $M=(v,f)$ ($f$ is the face list), and a plane $\Pi$. 
    \OUTPUT Two meshes $M_1=(v_1,f_1)$ and $M_2=(v_2,f_2)$, result of $M$ 
    getting cut by $\Pi$  \\
    Evaluate (using GA) and order the intersection points of $\Pi$ with each face of $M$. \\
    Evaluate the weights and bone indices that influence these points.
    \label{step:weights_cutting}\\
    Re-triangulate the faces that are cut using the intersection points.\\
    Separate faces in $f_1$ and $f_2$, depending on which side of the 
    plane they lie. \\
    From $f_1$ and $f_2$, construct $M_1$ and $M_2$. 
  \end{algorithmic}
\end{algorithm}


\subsubsection{Tearing Algorithm} 
\label{ssub:tearing_module}

\begin{figure}[hbt]
\centering
\subfloat[]{\includegraphics[width=0.25\textwidth]{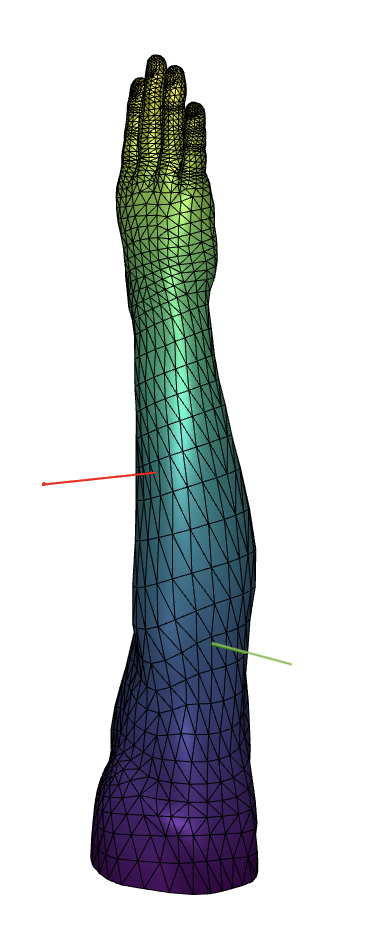}}
\hspace*{10mm}
\subfloat[]{\includegraphics[width=0.25\textwidth]{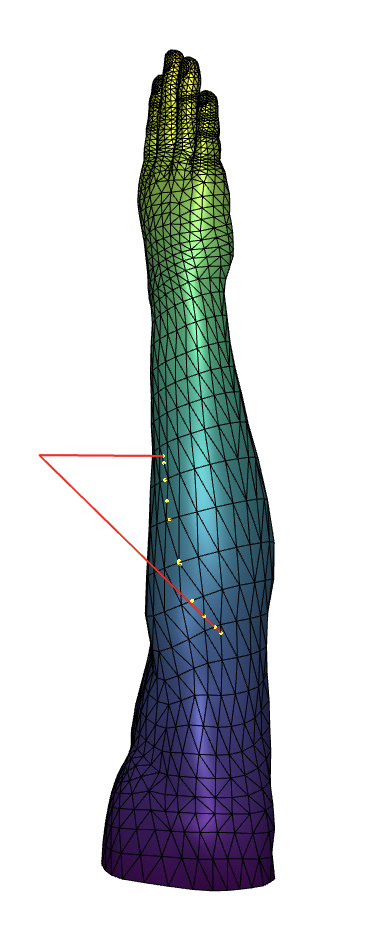}}
\hspace*{10mm}
\subfloat[]{\includegraphics[width=0.25\textwidth]{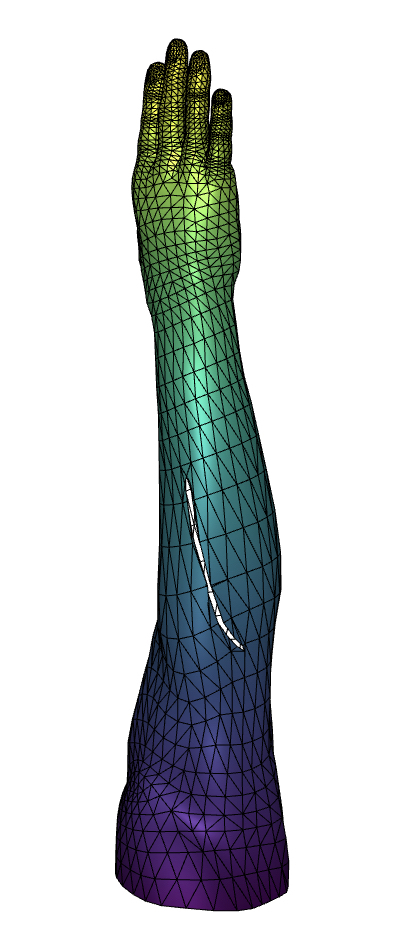}}
\caption{Tearing module intermediate steps. 
(A) The original animated model and the scalpel's position at two 
consecutive time steps.
(B) The plane defined by the scalpels
(depicted as a red triangle) intersects the skin in the yellow points.
(C) The intermediate points are used in the re-triangulation, and 
are «pushed» away from the cutting plane to form an open tear.}
\label{fig:tearing_module}
\end{figure}

\begin{figure}[bt!]
\centering
\subfloat[]{\includegraphics[width=0.2\textwidth]{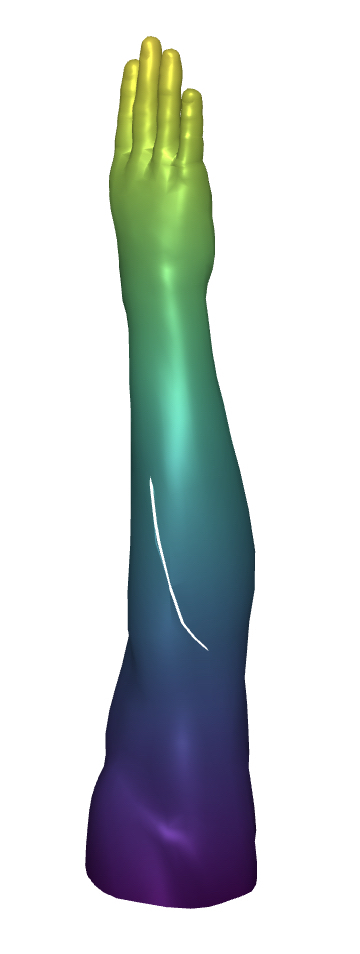}}
\hspace*{0mm}
\subfloat[]{\includegraphics[width=0.16\textwidth]{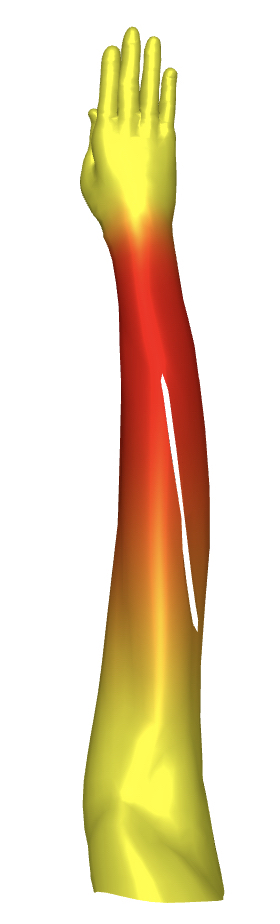}}
\hspace*{0mm}
\subfloat[]{\includegraphics[width=0.3\textwidth]{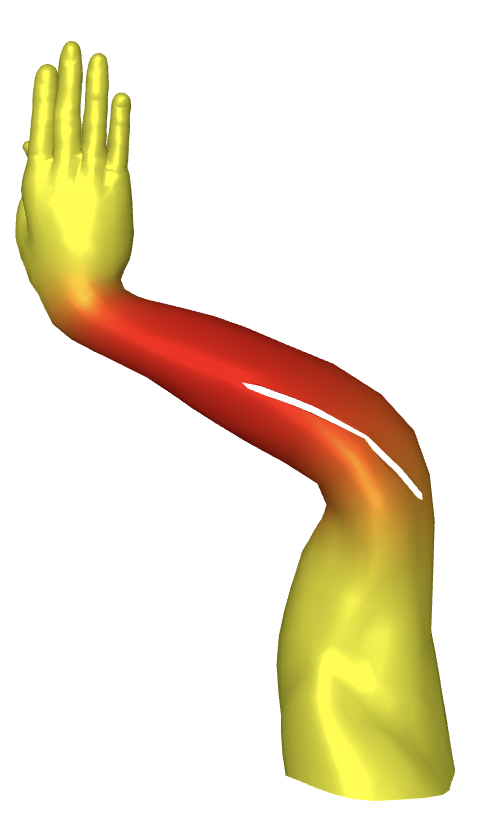}}
\hspace*{0mm}
\subfloat[]{\includegraphics[width=0.17\textwidth]{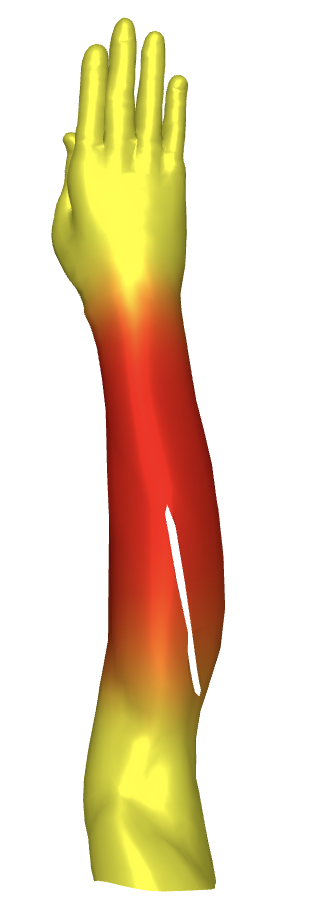}}
\caption{Deformation of a torn model. 
(A) The original model after applying the tear.
(B) Two rotations are applied to the torn model, 
one at elbow joint around $y$-axis by $\rad{-1}$, and another at
wrist joint around $y$-axis by $\rad{1}$.
(C) A dilation of scale 1.5 is applied to the torn model,
at elbow joint.
(D) A translation is applied to the torn model at elbow joint with 
translation vector $(18,0,0)$. In all cases, 
minor artifacts only arise, despite the great magnitude of the 
applied deformations. In (B),(C) and (D), vertices are colored 
depending on the influence of elbow joint which is mostly deformed. In (A), vertices are colored based on their $z$ coordinate.}
\label{fig:tearing_module_arm}
\end{figure}

  The purpose of this module is to enable partial cuts on the skinned
  model, in contrast with the cutting module where the cut is, in 
  a sense, complete. The importance of this module derives from the
  fact that most of the surgical incisions are partial cuts and 
  therefore they are worth replicating in the context of a virtual 
  surgery. Towards that direction, our work involves an algorithm 
  that both tears a skinned model and also enables animation of the
  final mesh (see Figures~\ref{fig:tearing_module} and \ref{fig:tearing_module_arm}). 

  To understand the philosophy behind the design of the tearing 
  algorithm that is described below, one must comprehend the 
  differences between cutting and tearing. In tearing, the movement 
  of a scalpel defines the tear rather than a single plane. 
  To capture such a tear in geometric terms, we have to take into
  consideration the location of the scalpel in either a continuous way
  (e.g., record the trail of both endpoints of the scalpel in terms 
  of time) or a discrete way (e.g., know the position of the scalpel 
  at certain times $t_i$). For VR purposes, the latter way is 
  preferred as it yields results with better fps, since input is
  hard to be monitored and logged continuously in a naive way. 
  For these reasons, our implementation requires the scalpel position 
  to be known for certain $t_i$. 

  The proposed tearing algorithm is summarized in   Algorithm~\ref{alg:tearing}. A description of how we tackle the weight evaluation in step~\ref{step:weights_tearing} is found in Section~\ref{sub:implementation_details}.

  \begin{algorithm}[!ht]
  \caption{Tearing Algorithm}
  \label{alg:tearing}
  \begin{algorithmic}[1]
    \INPUT Triangulated Mesh $M=(v,f)$, and scalpel position 
    at time steps $t_i$ and $t_{i+1}$
    \REQUIRE Scalpel properly intersects $M$ at these time steps
    \OUTPUT The mesh $M_t=(v_t,f_t)$ resulting from $M$ getting torn
    by the scalpel\\
    Determine the intersection points $S_i$ and 
    $S_{i+1}$ of $M$ with the scalpel at time step $t_i$ 
    and $t_{i+1}$ respectively. \\
    Determine the plane $\Pi$, containing 
    $S_i$ and the endpoints of scalpel at time $t_{i+1}$. Small time steps guarantee that $\Pi$ is well-defined. \\
    Evaluate the intersection points $Q_j$ of $\Pi$ and $M$, s.t. 
    the points $S_i$,$Q_0$,$Q_1$,$\ldots$,$Q_m$,$S_{i+1}$ 
    appear in this order on $\Pi$ when traversing the 
    skin from $S_i$ to $S_{i+1}$.\\
    Assign weights to points $S_i$, $S_{i+1}$ and all $Q_j$. \label{step:weights_tearing}\\
    Re-triangulate the torn mesh, duplicating $Q_j$ vertices.\\
    Move the two copies of $Q_j$ away from each other to 
    create a visible tear (optional).\label{step:open_tearing}
  \end{algorithmic}
  \end{algorithm}

  Our major assumption is that all intermediate intersection 
  points lie on this plane, which is equivalent to the assume 
  that the tearing curve is smooth, given that $t_i$ and $t_{i+1}$ 
  are close enough. In our implementation, during 
  step~\ref{step:open_tearing}, the intermediate torn points are  
  moved parallel to the direction of the normal of the plane 
  $\Pi$ and away from it, to replicate the opening of a cut 
  human tissue.
  

\subsubsection{Drilling Algorithm} 
\label{ssub:drilling_module}

The usage of Virtual Reality by surgeons and their need to drill 
holes in a simulated 3D model motivated the creation of the 
drilling module. Given a triangulated mesh and finite cylindrical drill, we would 
like to evaluate the mesh that corresponds to the drilled model. 

Designing the drilling predicate was more intriguing, compared to the 
respective cutting and tearing algorithms, as multiple ideas turned out to be
inadequate. The initial idea of substituting 
the cylinder with a prism of $n$-surfaces, for some suitable $n$, looked 
promising enough, as it would enable using drill as a special case of tear. 
However, one would have to provide an easy way to determine an $n$ that 
would be sufficiently large to produce a smooth hole-like effect in the 
outcome mesh. On the other hand, choosing an arbitrary large $n$ would 
result in many surfaces and therefore many costly tear operations had 
to be performed, hindering our chances of a real-time implementation. 
The prismatic approach also yielded the question of how to choose the 
position the edges of the prism such that the intersection points of the 
prism and the mesh would be re-triangulated in a clever and robust way. 
Of course, if the edges of the prism were selected such that they 
intersected the mesh's faces only on their boundaries, the re-triangulation 
would be more efficient and not produce a lot of slither faces. However, 
if we had to decide the optimal prism, that would be equivalent to 
specify the intersection points of all mesh edges with our initial 
cylinder, which is the idea behind our proposed algorithm.

In the core of our drill module lies a point-versus-cylinder predicate 
that allows us to determine the intersection point of every edge of 
the given mesh with the cylindrical drill. Since the drill is described by 
its radius $r$ and two endpoints $A$ (the ``tip'' of the drill) and $B$ 
that define its axis, we can easily determine the plane~$\Pi$ that is perpendicular to its axis and goes through~$B$. Given an edge $e$ defined by the vertices $v_i$ and $v_j$ of the mesh, we first determine if any of these two vertices lie inside the semi-finite cylinder 
(we ignore the existence of $A$ for now and consider that the cylinder is only 
bounded by~$\Pi$ and goes indefinitely towards the direction of $A$). 
To accomplish such task for the vertex $v\in\{v_i,v_j\}$, we project 
it to the plane $\Pi$ and compare the distance of the projected point $P(v)$
and $B$ with $r$; if it is smaller (respectively larger) then $v$ lies 
strictly inside (resp. outside) the cylinder. In the case of equality, 
the vertex $v$ lies on the cylinder. 

If the vertices $v_i$ and $v_j$ lie on different sides with respect to 
the cylinder, then we first evaluate the intersection point of the edge
defined by $\mu:=P(v_i)$ and $\nu:=P(v_j)$ with the sphere centered at $B$ with radius
$r$. The coordinates of $\mu$ and $\nu$ can be explicitly determined as they are the projections of the points $v_i$ and $v_j$ respectively on the plane
going through $B$ with normal $\vec{n} = \vec{AB}/||\vec{AB}||$. Therefore, 
the projected image of $\vec{v}\in\{v_i,v_j\}$ on the plane is $\vec{v}-\langle\vec{n},\vec{v}-\vec{OB}\rangle \vec{n}$, where 
$\langle\cdot,\cdot\rangle$ denotes the classic inner product.

Since every point on the projected edge is of the form $\alpha\mu+(1-\alpha)\nu$ for some $\alpha \in [0,1]$, and the edge is intersected by the sphere - as it is intersected by the cylinder- there exists an $\alpha$ that corresponds to the intersection point. For this $\alpha$, the point $\xi:=\alpha \mu+(1-\alpha \nu$ must have exact distance from~$B$ equal to $r$. 
If $d(\cdot,\cdot)$ denotes the Euclidean distance, solving the equation 
$d^2(\xi,B)=r^2$ in terms of $\alpha$ yields that $\alpha$ is a root of the 
quadratic equation $K\alpha^2+L\alpha+N=0$, where 
\begin{align}
K &= (x_\mu-x_\nu)^2 + (y_\mu-y_\nu)^2 + (z_\mu-z_\nu)^2 \neq 0,   \\
L &= 2(x_\mu-x_\nu)(x_\nu-x_B)+2(y_\mu-y_\nu)(y_\nu-y_B)\nonumber \\
& \quad +2(z_\mu-z_\nu)(z_\nu-z_B), \\
N &= (x_\nu-x_B)^2+(y_\nu-y_B)^2+(z_\nu-z_B)^2-r^2,
\end{align}
and $\xi = (x_\xi, y_\xi,z_\xi )$, for $\xi\in\{\mu,\nu,B\}$. 

\begin{figure}[tb!]
\centering
\subfloat[]{\includegraphics[width=0.18\textwidth]{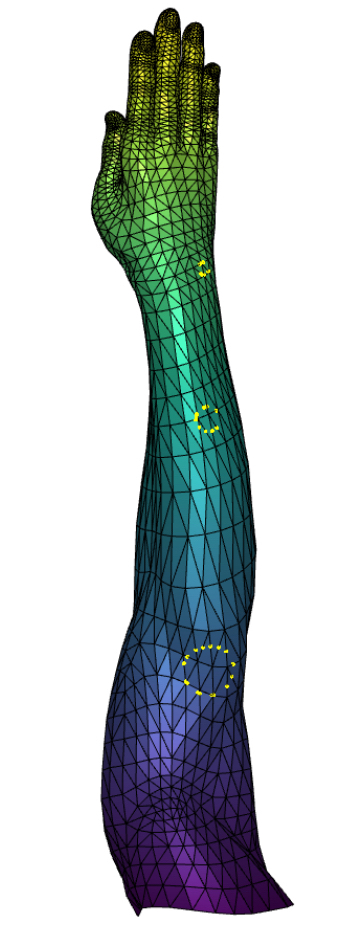}}
\hspace*{10mm}
\subfloat[]{\includegraphics[width=0.14\textwidth]{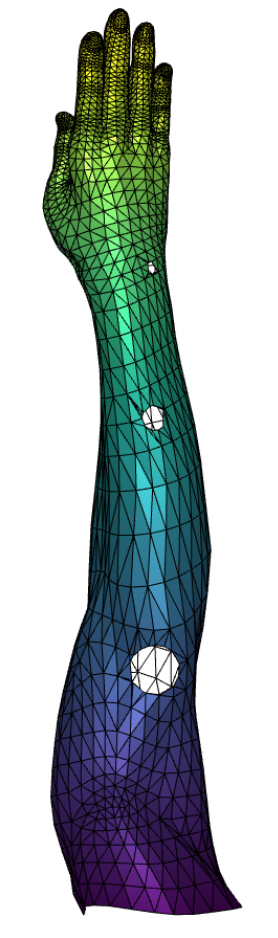}}
\hspace*{10mm}
\subfloat[]{\includegraphics[width=0.27\textwidth]{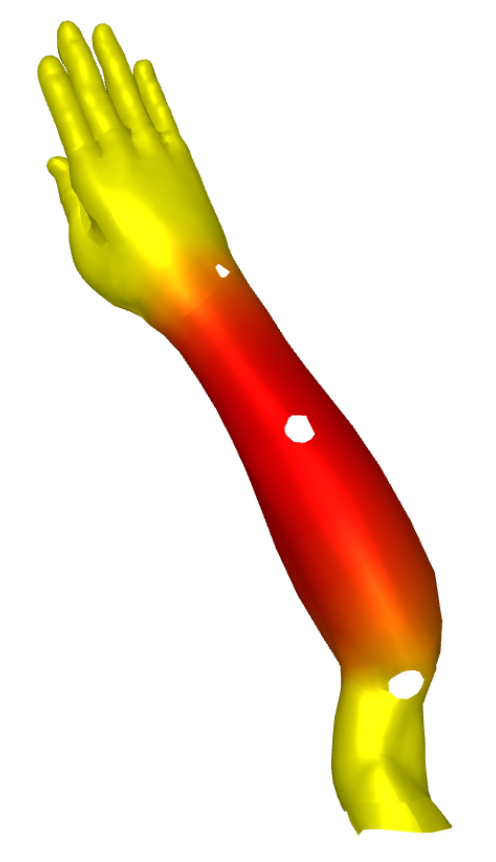}}
\caption{Drilling module intermediate steps. 
(A) The drill intersects the model skin in the yellow points.
(B) The intersection points are used in the re-triangulation.
(C) The elbow joint of the drilled model is translated by $(1,1,1)$, rotated 
by $\rad{0.3}$ around all 3 axis and then dilated by a factor of~2. The weight function ensures that minimal to no artifacts arise in the drilled area despite the deformation.}
\label{fig:drilling_module}
\end{figure}

Therefore, we conclude that $\alpha$ is the only root of the quadratic that belongs in $[0,1]$. Since for this $\alpha$, the point $\alpha P(v_i)+(1-\alpha)P(v_j)$ is the intersection of the 
cylinder with the projected edge, the point $\alpha v_i+(1-\alpha)v_j$ is a good approximation 
of the intersection point of cylinder with the original edge.

Except of the basic edge-cylinder intersection where the endpoints $v_i$
and $v_j$ of the edge lie on different sides with respect to the 
cylinder, another two cases have to be taken into consideration. 
It is possible that both 
endpoints lie outside of the cylinder but the edge intersects the 
cylinder in two points or is tangent to the cylinder in one point. 
These cases are equivalent to both $P(v_i)$ and $P(v_j)$ lying outside the 
spheres centered at $B$ with radius~$r$ and the quadratic 
$K\alpha^2+L\alpha+N=0$ has one or two root(s) $\alpha\in [0,1]$. As before, 
the intersection point(s) is(are) approximated by 
$\alpha v_i+(1-\alpha v_j$ for these $\alpha$. 

After evaluating the intersection points of the drill with the model 
and since all of them lie on some edge of the original mesh, a robust and 
efficient triangulation can be easily applied. If the number of intersection points 
is below some threshold, e.g., $6$, we can perform a ``split'' operation 
on all affected faces and drill again. To split a triangular face one 
may connect the middle points of all edges and therefore create four 
smaller sub-triangles similar to the original. This operation will create a 
more ``dense'' triangulation in the specific part of the model, 
resulting in more intersection points with the drill and hopefully in 
a more realistic result. Although generating more intersection points when needed is not a difficult task, we have to take into consideration that 
it has to be done in a clever way so as not to hinder the re-triangulation 
process in terms of performance or implementation complexity. 

The results of our drilling module when applied to our arm model are 
demonstrated in Figure~\ref{fig:drilling_module}. As in the precious modules, we can assign weights to the newly introduced intersection points, 
allowing re-deformation of the drilled model. 
The outline of the proposed drilling algorithm is described in 
Algorithm~\ref{alg:drilling}. A summary of how we address 
the weight evaluation in step~\ref{step:weights_drilling}
is found in Section~\ref{sub:implementation_details}. 

\begin{algorithm}[hb]
\caption{Drilling Algorithm}
\label{alg:drilling}
\begin{algorithmic}[1]
  \INPUT Triangulated Mesh $M=(v,f)$ and drill position via its 
  endpoints $A$ (``tip'' of the drill),$B$ and radius $r$.
  \REQUIRE Drill properly intersects $M$
  \OUTPUT The mesh $M'=(v',f')$ resulting from $M$ being drilled\\
  Let $\Pi$ denote the plane perpendicular to the drill axis going through
  the endpoint $B$. \\
  Determine the faces of $M$ that are pierced by the 
  drill and run a BFS algorithm that checks, for this and all neighboring 
  faces, if at least one of its three vertices, when projected to $\Pi$,
  has distance from~$B$ less than~$r$, i.e., if it lies within the drill. 
  Mark such a face as ``affected''.\\
  For all affected faces, determine in the plane $\Pi$ the intersection 
  points of the drill and the projected face and then down-project them
  to the original mesh.\\  
  Assign weights to the intersection points of the original mesh.\label{step:weights_drilling}\\
  Re-triangulate the drilled mesh by replacing the affected faces by 
  appropriate ones.
\end{algorithmic}
\end{algorithm}
  

\subsection{Implementation Details and Performance Remarks} 
\label{sub:implementation_details}

The main framework used for skinning and animation with the use of multivectors is Python's PyAssimp\footnote[1]{{PyAssimp} Homepage: \url{https://pypi.org/project/pyassimp/}} and Clifford\footnote[2]{{Clifford} Homepage: \url{https://clifford.readthedocs.io/}} package 
for the evaluation of the vertices and the Meshplot package for rendering the 
model. The use of Python language was preferred for a more user and presentation-friendly experience; for a more robust and efficient implementation C++ would be advised. 

An instance of a class called v\_w is used to store 
for each vertex a list of up to 4 bones that influence it along with 
the corresponding influence factors. The node tree is then traversed 
and all information regarding rotation, translation 
and dilation are translated to multivectors \cite{DietmarFoundations,DorstBook} and also stored in the instance for convenience. 
In order to evaluate the final position of the vertices, all that
is left is to to evaluate the sum in equation~\eqref{eq:cga_formula} for all 
vertices and down project it to $\mathbb{R}^3$, for each vertex. 
There are two possible ways of achieving this task. The first way is 
to evaluate the sum and then 
down project the final result to obtain each vertex in Euclidean form. 
The second way is to down project each term and then add them to get
the final result. Although not obvious, the second method yields 
faster results since the addition of 4 multivectors (32-dimensional arrays)
and one down-projection is slower than down-projecting (up to) 4 
multivectors and adding 4 euclidean vectors of dimension 3.

A final implementation detail regards the weight evaluation for 
newly added vertices in the cutting and tearing modules. In the
former module, such vertices necessarily lie on an edge of the 
original mesh, whose endpoints both lie on different sides of the 
cutting plane. 
Another method is the one used in the tearing 
module where the intersection point can also lie inside a face. 
Assuming the point $X$ lie somewhere on the face $ABC$, we can 
explicitly write $OX = p OA + q OB + r OC$ for 
some $a,b,c\in [0,1 ]$ such that $p+q+r=1$. The tuple $(p,q,r)$ 
is called the \emph{barycentric coordinate} of $X$ with respect to the 
triangle $ABC$. Each of the vertices $A,B,C$ are (usually) 
influenced by up to 4 bones, so let us consider that they are all 
influenced by a set of $N(\leq 12)$ vertices, where the bones 
beside the original 4 have weight 0. Let $w_A,w_B,w_C,w_X$ denote the 
vectors containing the $N$ weights that correspond to 
vertices $A,B,C$ and $X$ respectively, for the same ordering of the 
$N$ involved bones. To determine $w_X$, we first evaluate 
$w = p w_A + q w_B + r w_C$ and consider two cases. 
If $w$ contains up to 4 non-zero weights, then $w_X = w$. Otherwise, 
since each vertex can be influenced by up to 4 bones, we keep the 
4 greater values of $w$, set the others to zero, and normalize the 
vector so that the sum of the 4 values add to 1; the final result
is returned as $w_X$. We denote this weight as \emph{weight of $X$ 
via barycentric coordinates}. Variations of this technique can be 
applied in both modules to prioritize or neglect influences on 
vertices lying on a specific side of the cutting plane. Different 
variations of the weight function allows for less artifacts \cite{wareham2008bone}, 
depending on the model and the deformation subsequent to the 
cutting/tearing.

\textbf{Performance:} Running the cutting algorithm in the arm model 
(5037 faces, 3069 vertices)
took for a simple cylinder model a total of $\ms{898}$:  $\ms{42}$ for vertex separation,
$\ms{757}$ for re-triangulation of the 92 intersection points, $\ms{87}$ to split 
faces in two meshes and $\ms{12}$ to update the weights. To cut the arm model, 
it took $\ms{4666}$ as shown in Table~\ref{tab:cutting_running_time}, where 
most time ($\ms{2205}$) was spent on the evaluation and triangulation of 
the intersection points of the cutting plane and the model. The 
offline pre-processing time of the model, i.e., the time required 
to translate the model skin or animation data from Euclidean 
coordinates or matrices respectively to multivector form is not taken 
into account in the above measurements. 

\begin{table}[tb]
 \centering
 \begin{tabular}{|c || c | c |} 
 \hline
 & Time Spent Using & Time Spent Using \\
 Subroutine & Euclidean Tools & GA Tools \\ [0.5ex] 
 \hline\hline
 Subroutine 1 & $\sec{0,036439578 +}$ & $\sec{0,072434584}$\\ 
 \hline
 Subroutine 2 & $\sec{2,050984303}$ & $\sec{2,205727577}$\\
 \hline
 Subroutine 3 & \multicolumn{2}{|c|}{$\sec{0,061544961}$} \\
 \hline
 Subroutine 4 & \multicolumn{2}{|c|}{$\sec{2,326937914}$} \\
 \hline\hline
 Cutting time & $\sec{4,475906756}$ & $\sec{4,666645036}$\\ [1ex] 
 \hline
\end{tabular}
\label{tab:cutting_running_time}
\caption{Running times of the four main subroutines of 
the cutting algorithm. In the 2nd column, point-versus-plane 
relative positions for subroutine 1 and segment-plane intersections
for subroutine 2 were determined using only Euclidean subpredicates. 
In the 3rd column, the same operations were carried using 
Geometric Algebra equivalent subpredicates. The subroutines 3 and 
4 are independent of the model data representation.
Subroutines: (1) Check vertices locations with respect to the cutting 
plane, (2) Detect which faces are intersected by the cutting plane, 
evaluate the intersection points and triangulate them, (3) Evaluate weights for the intersection points, (4) Split original model into submodels.}
\end{table}

Applying the tearing algorithm to the arm model took $\ms{2437}$ for the final output, 
for 34 intersection points. Most of this time ($\ms{2411}$) were needed just to determine which two faces were intersected by the scalpel. Tearing a simple cylinder model (758 faces, 634 vertices) took $\ms{362}$ for 17 intersection points. Again, most time ($\ms{331}$) was spend for the scalpel intersection. 

The drilling algorithm for the arm model takes on average $\ms{274}$ for a hole consisting of 17 intersection points on our arm model. For a hole of the same diameter consisting of 20 intersection points, the algorithm requires $\ms{319}$ to return the final outcome whereas,
the running time grows to $\ms{595}$ when the diameter is increased 
from~2 to~3 and the intersection points become~33. As a rule of thumb, 
there is an average running time of $\ms{16-18}$  per intersection point. 

These running times, produced in a MacbookPro with a  2,6 GHz 6-Core Intel Core i7 processor, can be greatly improved as our current 
unoptimized  CPU-based Python implementation has to thoroughly 
search all faces for cuts/tears. Multivector operations are performed 
by the Clifford python package which, in some cases, allows some 
parallelization. However, since python list comprehensions and functions 
of multiple types of inputs are involved in 
our implementation, we could not fully parallelize our algorithm and python 
performed most operations in a single thread. 
A GPU implementation optimized 
for multivector operations would allow the comparison of our 
proposed method with the current state-of-the-art methods, 
which however do not allow further deformation 
of the model. 
The running times of our algorithms indicate that
there is only a small percentage of performance load added when 
using Geometric Algebra representation forms instead of Euclidean 
ones to perform cuts/tears and drill holes. 

\section{Conclusions and Future Work} 
\label{sec:conclusions_and_future_work}

This work describes a novel way to perform model animation 
and deformation as well as cutting, tearing and drilling under 
a single geometric framework in
Conformal Geometric Algebra. We focus towards a pure 
geometric-based implementation that can be applied to 
rigged models even in low-spec VR headsets and ultimately enable 
real-time operations such as the ones presented here. 
Our current results were obtained using python but,
since our goal is to have a full implementation in real-time virtual 
reality simulation, we will inevitably have to use more suitable 
programming languages and platforms such as C\#/C++ and 
Unity/Unreal Engines.
It is our intention to use recently developed acceleration techniques \cite{Hadfield:2019fm} and parallel processing to further optimize our 
algorithms and further decrease running times. We currently redesign 
parts of the algorithm to allow parallelization in all functions where 
this is applicable. Finally, we intend to 
combine our modules in conjunction with a physics engine to obtain 
a realistic opening effect, e.g., after the user performs a tear, 
without the need to pre-record it and therefore lift the 
limitation of only predefined, physics-based,plausible cuts,tears or drills. 


\section{Acknowledgements} 
\label{sec:acknowledgements}
We would like to cordially thank the anonymous paper reviewers as 
well as the handling editor for the constructive and helpful comments.

\bibliographystyle{spmpsci}
\bibliography{references2}

\end{document}